
\input phyzzx


\def\IMPA#1{{\sl Int. J. Mod. Phys. {\bf A#1}}}
\def\IMPB#1{{\sl Int. J. Mod. Phys. {\bf B#1}}}

\def\JMP#1{{\sl J.\ Math. \ Phys.\ {\bf #1}}}

\def\JPA#1{{\sl J.\ Phys.\ {\bf A#1}}}

\def\NPB#1{{\sl Nucl.\ Phys.\ {\bf B#1}}}

\def\PRL#1{{\sl Phys.\ Rev.\ Lett.\ {\bf #1}}}

\def\PTP#1{{\sl Prog.\ Theor.\ Phys.\ {\bf #1}}}


%
\def\nxl{\hfill\break}



\def\t{\theta}

\def\b{\beta}
\def\g{\gamma}
\def\l{\lambda}

\def\e{\epsilon}
\def\o{\over}

\def\s{\sigma}

\def\Th{\Theta}

\def\sub#1#2{{{#1^{\vphantom{0000}}}_{#2}}}
\def\frac#1#2{{\textstyle{
 #1 \over #2 }}}                            


\def\ZZ{{\rm Z \!\! Z}}                       
\def\1{{\rm 1 \!\!\, l}}                        
%

%
%


\hyphenation{Di-par-ti-men-to}
\hyphenation{na-me-ly}
\hyphenation{al-go-ri-thm}
\hyphenation{pre-ci-sion}
\hyphenation{cal-cu-la-ted}

%

%

\Pubnum={$\rm PAR\; LPTHE\; 92/30$}
\date={}
\titlepage
\title{
SIMPLE APPROACH TO THERMAL BETHE ANSATZ}
\author{ H.J. de Vega }
\address{ Laboratoire de Physique Th\'eorique et Hautes Energies
     \foot{Laboratoire Associ\'e au CNRS UA 280}, Paris
     \foot{mail address: \nxl
           L.P.T.H.E., Tour 16 $1^{\rm er}$ \'etage, Universit\'e Paris VI,\nxl
           4 Place Jussieu, 75252, Paris cedex 05, France }}
\author{Lecture delivered at the XXI International Conference on
Differential Geometry Methods in Theoretical Physics, Tianjin, P. R.
of CHINA, June 5-9, 1992}
\endpage
\title{
SIMPLE APPROACH TO THERMAL BETHE ANSATZ}

\author{ H.J. de Vega }
\address{ Laboratoire de Physique Th\'eorique et Hautes Energies
     \foot{Laboratoire Associ\'e au CNRS UA 280}, Paris
     \foot{mail address: \nxl
           L.P.T.H.E., Tour 16 $1^{\rm er}$ \'etage, Universit\'e Paris VI,\nxl
           4 Place Jussieu, 75252, Paris cedex 05, France }}

\vfil
\abstract
We report on a new approach to the calculation of thermodynamic functions
for crossing-invariant models solvable by Bethe Ansatz. In the case of the
XXZ Heisenberg chain we derive, for arbitrary values of the anysotropy,
a {\bf single} non--linear integral equation
from which the free energy can be exactly calculated.These equations
are shown to be equivalent to an infinite set of algebraic equations
of Bethe type which provide alternatively the thermodinamics.
 The high--temperature expansion follows in a sistematic and
relatively simple way from our non-linear integral equations. For low
temperatures we obtain the correct central charge and predict the analytic
structure of the full expansion around $T=0$. Furthermore, we derive a
single non-linear integral equation describing the finite--size ground--state
energy of the Sine--Gordon quantum field theory.

\medskip

\vskip 1cm

\sequentialequations
\def\intf{\int_{-\infty}^{+\infty}\!}
\def\bt{{\tilde\beta}}
\def\phip{\phi^\prime}

\REF\yy{C.N. Yang and C.P. Yang, \JMP{10} (1969 1115.}
\REF\tak{M. Takahashi, \PTP{46} (1971) 401. \nxl
         M. Takahashi and M. Suzuki, \PTP{48} (1972) 2187.}
\REF\gaud{M. Gaudin, \PRL{26} (1971) 1302.}
\REF\nos{C. Destri and H. J. de Vega,LPTHE Paris preprint, 92-11 (to
appear in \PRL ).}
\REF\noss{C. Destri and H. J. de Vega, in preparation.}
\REF\baxt{R.J. Baxter, ``{\it Exactly solved models in Statistical Mechanics"},
          Academic Press, 1982. }
\REF\japs{J. Suzuki, Y. Akutsu and M. Wadati, {\sl J. Phys. Soc. Jpn.} {\bf 59}
          (1990) 2667.}
\REF\ddev{C. Destri and H.J. de Vega, \NPB{290} (1987) 363; \JPA{22} (1989)
          1329.}
\REF\dev{H.J. de Vega, \IMPA{} (1989) 2371; \IMPB{} (1990) 735.}
\REF\devw{H.J. de Vega and F. Woynarovich, \NPB{251} (1985) 439.}
\REF\devk{H.J. de Vega and M. Karowski, \NPB{280} (1987) 225.}
\REF\aus{A. Klumper, M.T. Batchelor and P.A. Pearce, \JPA{24} (1991) 3111.}
\REF\dedev{C. Destri and H.J. de Vega, \NPB{258} (1991) 251.}
\REF\albz{Al.B. Zamolodchikov, \NPB{342} (1990) 695.}

The computation of thermodynamical functions for integrable models started with
the seminal works of C.N. Yang and C.P. Yang [\yy], of  M. Takahashi  [\tak]
and M. Gaudin [\gaud]. In Refs. [\tak,\gaud] the free energy of the Heisenberg
spin chain is written in terms of the solution of an infinite set of coupled
nonlinear integral equations, derived on the basis of the so--called ``string
hypothesis". In ref.[\nos], we proposed a simpler way to solve the
thermodynamics
by means of a single, rigorously derived, nonlinear integral equation.
This method can be applied to a wide class of models solvable by Bethe Ansatz
(see ref.[\noss]).

The XXZ hamiltonian for a periodic chain with $2L$ sites (we assume
periodic boundary conditions but generalization to other b.c. is possible)
takes the form
$$
      H_{XXZ}=-J\sum_{n=1}^{2L} \left[
      \s^x_n\s^x_{n+1} + \s^y_n\s^y_{n+1} -\cos\g\,(\s^z_n\s^z_{n+1}+1)
                                      \right]                   \eqn\hxxz
$$
The anisotropy parameter $\g$ is assumed here to lay in the real interval
$(0,\,\pi)$ characteristic of the gapless regime. $H_{XXZ}$ is related to
the diagonal--to--diagonal transfer matrix $T_L(\t)$ of the six--vertex model
by $$
   T_L(\t)\buildrel {\t\to0} \over =
1-{\t\o{2J\sin\g}}H_{XXZ} + O(\t^2) \eqn\ttoh
$$
where $$
 T_L(\t)=R_{12}R_{34}\ldots R_{2L-1\,2L}R_{23}R_{45}\ldots R_{2L\,1}  \eqn\tran
$$ and
$$\eqalign{  R_{nm} =&{{a+c}\o2}+{{a-c}\o2}\s^z_n\s^z_m
                     +{b\o2}(\s^x_n\s^x_{n+1} + \s^y_n\s^y_{n+1})      \cr
    a =&{{\sin(\g-\t)}\o{\sin\g}}\;,\quad
          b ={{\sin\t}\o{\sin\g}}\;,\quad c=1           \cr}       \eqn\rma
$$
$\t$ is known as spectral parameter while $a$, $b$, and $c$ are the
conventional six--vertex Boltzmann weights.

The free energy per site is defined as usual
$$
      f(\b)=-{1\o\b}\lim_{L\to\infty}{1\o{2L}}\log\left[
            {\rm Tr}\left(e^{-\b H_{XXZ}}\right)\right]       \eqn\free
$$
and from eq. \ttoh\ we read
$$
   e^{-\b H_{XXZ}}= \lim_{N\to\infty} \left[T_L(2\bt/N)\right]^N
$$
where $\bt=\b J\sin\g$. Hence we can write
$$
     f(\b)=-{1\o\b}\lim_{L\to\infty}{1\o{2L}}\lim_{N\to\infty}
           \log Z_{LN}(2\bt/N)                           \eqn\part
$$
where $Z_{LN}(\t)\equiv {\rm Tr\,}[T_L(\t)]^N$ is the six--vertex
partition function on a periodic diagonal lattice with $L\times N$ sites.
The two limits in eq. \part\ cannot be interchanged since the degeneracy
of $T_L(0)=1$, that is $2^{2L}$, is strongly $L-$dependent. However, the
crossing invariance of the six--vertex $R-$matrix
implies that under a rotation by $\pi/2$ of the entire lattice plus
the substitution $\t\to\g-\t$, the numerical value of the partition
function does not change (see $e.g.$ ref.\baxt).
Therefore $Z_{LN}(\t)=Z_{NL}(\g-\t)$ and
$$
     f(\b)=-{1\o\b}\lim_{L\to\infty}{1\o{2L}}\lim_{N\to\infty}
           \log Z_{NL}(\g-2\bt/N)                           \eqn\crossed
$$
where $Z_{NL}(\t)\equiv {\rm Tr\,}[T_N(\t)]^L$. Now, as the well--known
Bethe Ansatz solution tells us, $T_N(\g)$ has a non--degenerate largest
eigenvalue for any finite $N$. Then the two limits in eq. \crossed\ commute
and one finds [\japs]
$$
    -2\b f(\b)=\lim_{N\to\infty}\log \Lambda_N^{max}(\g-2\bt/N) \eqn\eigmax
$$
where $\Lambda_N^{max}(\t)$ denotes the largest eigenvalue of $T_N(\t)$.
Using the results of ref.[\ddev] for the eigenvalues of the diagonal--to--
diagonal transfer matrix, the free energy becomes
$$
     f(\b)={1\o{2\b}}\lim_{N\to\infty}\left[E_N(\b)-
            2N\log{{\sin(2\bt/N)}\o{\sin\g}}\right]             \eqn\subtra
$$
where $$\eqalign{
   E_N(\b) &=i\sum_{j=1}^N \e(\l_j,\b)             \cr
   \e(\l,\b)&=\phi(\l_j+i\g/2,\g/2-\bt/N) -\phi(\l_j-i\g/2,\g/2-\bt/N)\cr
   \phi(\l,x)&\equiv i\log{{\sinh(ix+\l)}\o{\sinh(ix-\l)}} \; \qquad
                               (\phi(0,x)=0)     \cr}       \eqn\defi
$$
The real numbers $\l_1,\ldots,\l_N$ are the roots of the Bethe Ansatz
equations $$
   N\left[\phi(\l_j,\bt/N)+\phi(\l_j,\g-\bt/N)\right]=
      \sum_{k=1}^N \phi(\l_j-\l_k,\g) +\pi (2j-N-1)                  \eqn\bae
$$
associated to the largest eigenvalue $\Lambda_N^{max}(\g-2\bt/N)$ (we assumed
$N$ even). We define as usual the {\it counting function} [\dev]
$$
     \sub zN(\l)={1\o{2\pi}}\left[\phi(\l,\bt/N)+\phi(\l,\g-\bt/N)-
     {1\o N} \sum_{k=1}^N \phi(\l-\l_k,\g)\right]                   \eqn\count
$$
which by construction enjoys the properties
$\sub zN(\l)=-\sub zN(-\l)=\overline{\sub zN({\bar\l})}$ and
$$
             \sub zN(\l_j)=j-(N+1)/2                             \eqn\prop
$$
Our aim is to derive an integral equation for $\sub zN(\l)$.
Let us start by assuming that the root distribution becomes the continuus
density $\sub\s{c}(\l)$ (normalized to 1) in
the limit $N\to\infty$. Then it
could be identified with the derivative of $\sub zN(\l)$ and at the same
time used to replace the discrete sum in eq. \count\ with an integration.
For large but finite $N$ we whould therefore obtain the following
approximate linear integral equation for $\sub\s{c}(\l)$
$$
    2\pi\s_c(\l)=\phip(\l,\bt/N)+\phip(\l,\g-\bt/N) -
    \intf d\mu\,\phip(\l-\mu,\g)\s_c(\mu)                   \eqn\linear
$$
Fourier transforming leads to the explicit solution
$$
      \s_c(\l) = z_c^\prime(\l) \; ,\qquad
      z_c(\l)= {1\o\pi}\arctan\left[
      {{\sinh(\pi\l/\g)}\o{\sin(\pi\bt/\g N)}}\right]          \eqn\zetac
$$
We see that $\s_c(\l)$ is strongly peaked at $\l=0$ for large $N$, reflecting
the singularity which develops at the origin in the $N\to\infty$ limit of the
source term in eqs. \bae, \linear. The density picture correctly describes
the BA roots wherever $\s_c(\l)$ is of order 1. From eq. \zetac\ we then
find as validity interval
$$
                       |\l| \leq O(\sqrt{\b/N})                  \eqn\valid
$$
which shrinks to zero when $N\to\infty$. Roots $|\l_j|>O(\sqrt{\b/N})$ have a
spacing of order larger than $O(1/N)$ and cannot be descried by densities.
In particular, the roots with largest magnitudes have finite $N\to\infty$
limits spread by $O(1)$ intervals (we checked this fact numerically too).
Therefore, contrary to the usual situation [\dev], we must go beyond the
density description to obtain a bulk quantity like the free energy per site.
The sum in the last term of eq.\count\ can be written as a contour
integral
$$
\left\{1\o N\right\}\sum_{j=1}^N \phi(\l-\l_j,\g)=\int{{dz}\o{2\pi i N}}
\phi(\l-z,\g) {d\log[1 + \sub aN(z)]\o{dz}}	 		
	 \eqn\sumi
$$
where $\sub aN= \exp(2\pi iN\sub zN)$ and the contour in eq.\sumi\ encircles
counterclockwise all the real roots $\l _k,  k = 1,...,N$. Notice that
$\sub aN(\l_k) + 1 = 0,  k=1,...,N$. Inserting eq.\sumi\ in eq.\count\
and
integrating by parts yields
$$
\log{\sub
aN(\l))}=iN\left[\phi(\l,\bt/N)+\phi(\l,\g-\bt/N)\right]+i\intf
{dx\o\pi}\phip(\l-x,\g)\rm Im\log[1 + \sub aN(x+i0)]
							\eqn\segunda
$$
Convoluting eq.\segunda\ with the kernel
$$
			K^{-1}=\delta(\l)-p(\l)
$$
where$$
      p(\l)=\intf{{dk}\o{2\pi}}\,{\tilde p}(k)e^{ik\l}
           \; ,\quad     {\tilde p}(k)={{\sinh(\pi/2-\g)k}\o
              {2\sinh(\pi-\g)k/2 \,\cosh(\g k/2)}}                 \eqn\kernel
$$
we find
$$
    \log{{\sub aN(\l)}\o{a_c(\l)}}=\intf d\mu\, p(\l-\mu)
    \log{{1+\sub aN(\mu+i0)}\o{1+\overline{\sub aN(\mu+i0)}}}   \eqn\nonlin
$$
which is the sought nonlinear integral equation.(Here $a_c=\exp(2\pi iNz_c)$).
Notice that $\overline{\sub aN(\l)}=1/\sub aN({\bar\l})$.
We evaluate the sum in eq. \defi\ by a similar procedure, with the
result $$
    L_N(\b)\equiv E_N(\b)-E_c(\b)= {1\o\g}\intf d\l\,
    {{\sinh\pi\l/\g\,\cos\pi\bt/\g N}\o{\cosh2\pi\l/\g-\cos2\pi\bt/\g N}}
    \log{{1+\sub aN(\l+i0)}\o{1+\overline{\sub aN(\l+i0)}}}    \eqn\ener
$$
where $E_c(\b)$ is that part of the energy that follows from the root
density \zetac, that is
$$
    E_c(\b)=\intf d\l\, \e(\l,\g)\s_c(\l)= -2N\intf{{dk}\o k}\,
    {{\sinh(\pi-\g)k\,\sinh(\g-2\bt/N)k}\o{\sinh\pi k\cosh\g k}}   \eqn\econt
$$
If eqs. \nonlin\ and \ener\ are analytically continued to the axis
${\rm Im\,}\l=\pm\g/2$ they assume the same structure of those derived
in ref. [\aus] by different methods.

Inserting eqs. \ener\ and \econt\ in eq. \subtra\ yields
$$
      f(\b)= E_{XXZ}(\g) +\b^{-1}L(\b)                       \eqn\separ
$$
where $E_{XXZ}(\g)$ is the ground--state energy of \hxxz, namely
$$
     E_{XXZ}(\g)=2J\left[\cos\g-\sin\g\, \intf dk\,
       {{\sinh(\pi-\g)k}\o{\sinh\pi k\cosh\g k}}\right]            \eqn\exxz
$$
while $L(\b)$ is the $N\to\infty$ limit of \ener, that is
$$
        L(\b)={\rm Im}\intf d\l\,
        {{\log[1+a(\l+i0)]}\o{\g\sinh\pi(\l+i0)/\g}}             \eqn\correc
$$
Similarly, the function $a(\l)$ in this expression is the $N\to\infty$
limit of $\sub aN(\l)$. It is therefore the solution of the $N\to\infty$
limit of eq. \nonlin, that is
$$
        -i\log a(\l)= -{{2\pi\bt}\o{\g\sinh\pi\l/\g}}
        +2\intf d\mu\,p(\l-\mu)\,{\rm Im\,}\log[1+a(\mu+i0)] \eqn\nonlinear
$$
Thus the calculation of the free energy \free\ has been reduced to the
problem of solving the single nonlinear integral equation \nonlinear\ and
then evaluating the integral \correc.

We find from the $N=\infty$ limit of eq.\segunda\ an alternative form
for our eq. \nonlinear\
$$
  z(\l)=\bt q(\l) + {1\o{2\pi^2}}\,{\rm Im}\intf d\mu\, \phip(\l-\mu,\g)
                              \log\cos\pi z(\mu+i0)           \eqn\altform
$$
where $$
           q(\l)={1\o{\pi}}\left({{\sinh2\l}\o{\cosh2\l-\cos2\g}}-
                                       \coth\l \right)             \eqn\defq
$$
and $z(\l)\equiv \frac1{2\pi i}\log a(\l)$ is related to the original
counting function
$$
  z(\l)=\lim_{N\to\infty}\,N\left[\sub zN(\l)-\frac12{\rm sign\,}\l
        \right]     \; \quad (\l\; {\rm real})                    \eqn\relz
$$
By the residue theorem we then obtain
$$
     z(\l)=\bt q(\l)-{1\o{2\pi}}\sum_{j\in\ZZ}\left[\phi(\l-\xi_j,\g)
                              -\phi(\l,\g)\right]                \eqn\discre
$$
where the real numbers $\xi_j$, defined by $z(\xi_j)=j-1/2$, $j\in\ZZ$,
can be identified with the $N\to\infty$ limit of the original BA roots
$\l_1,\l_2,\ldots,\l_N$.The new algebraic BA equations
$z(\xi_j)=j-1/2$ embody all the information about the XXZ thermodynamics.  Eq.
\discre\ shows that $z(\l)$ has periodicity
$i\pi$ and has, as unique singularity on the real axis, a simple pole
at the origin with residue $-\bt/\pi$.
The new algebraic BA equations
$z(\xi_j)=j-1/2,j\in\ZZ$ embody all the information about the XXZ
thermodynamics.They form an infinite set of
Bethe Ansatz type algebraic equations equivalent to our non-linear
integral equations \nonlinear\ and \altform\ . We want to stress that
these roots are {\bf discrete} although we have already set $N =
\infty $. The root spacing is actually of order $T$. In the $T\to 0$
limit we recover the continuous distribution of roots associated with
the antiferromagnetic ground-state[\dev]. In terms of the roots
$\xi_k$, the free energy takes the form
$$
	f(\b)=J\cos\g-\b^{-1}\ln2+{1\o2\b}\sum_{k\in\ZZ}\log{([1-\cos
2\g]/[\cosh2\xi_j - \cos2\g])}
								\eqn\efetzi
$$
The roots $\xi_j$ have an accumulation point at $\xi = 0$. That is,
$\xi_j = -\bt/[\pi(j-{1/2}]$ for $j\to\infty$. For low temperatures,
the largest root, $\xi_1$, is of order ${(\g/\pi)}\ln\bt$.

Let us now study $f(\b)$ for high temperatures. When $\b$ is small
it is convenient to use the alternative form \altform\ plus the uniform
expansion$$
            z(\l)=\sum_{k=1}^\infty \bt^k\,b_k(\l)              \eqn\unif
$$
Since the residue of $z(\l)$ is linear in $\bt$ and $z(\l)$ is an odd function,
we have $$
          b_1(\l)\buildrel{\l\to0}\over\simeq -{1\o{\pi\l}}+O(\l) \;,\quad
          b_k(0)=0 \quad (k\ge 2)                              \eqn\bprop
$$
Inserting eq. \unif\ in eq. \altform\ we find, up to fourth order in $\bt$
$$\eqalign{
  b_1(\l) &=\vphantom{1\o{4\pi}}q(\l)\cr b_3(\l) &=\vphantom{1\o{4\pi}}0 \cr}
\qquad   \eqalign{
      b_2(\l) &={1\o{4\pi}}\phi^{\prime\prime}(\l,\g)\cr
      b_4(\l) &={1\o{6\pi}}\left({1\o3}-{1\o{{\rm sin}^2\,\g}}\right)
                \phi^{\prime\prime}(\l,\g) +{1\o{144\pi}}
                \phi^{\prime\prime\prime\prime}(\l,\g)\cr}     \eqn\fourth
$$
Then, from eqs. \separ\ and \correc,
$$
       f(\b)=-\b^{-1}\ln2 +J\cos\g-\b J^2\left(
        1+\frac12{\rm cos}^2\,\g\right)+\b^2J^3\cos\g+O(\b^3)
              \eqn\third
$$
which indeed agrees with the high $T$ expansion [\tak]. We want to remark
that eq. \altform\ generates the functions $b_k(\l)$ {\it recursively},
with easy integrations which involve only delta functions and derivatives
thereof. It is indeed a very efficent way to recover the high--temperature
expansion from the Bethe Ansatz solution [\noss].

We shall consider now the low temperature regime. When $\b\gg1$ eq. \nonlinear\
indicates that $z(\l)\sim\b$, so that $\log[1+a(\l)]\simeq 2\pi iz(\l)$,
at least as long as $|\l|\leq (\g/\pi)\log\b$. At dominant $\b\gg1$ order
eq. \nonlinear\ linearizes in the same way as it does for small $\b$.
Inserting then $z(\l)\simeq \bt q(\l)$ in eq. \correc, yields
$\lim_{\b\to\infty}L(\b)=0$. Therefore eq. \separ\ tells us that
$$
     f(\b)\buildrel {\b\to\infty} \over = E_{XXZ}(\g)+O(\b^{-2})\eqn\smallb
$$
The contributions of order $\b^{-2}$ and smaller come from values of $\l$
larger than $(\g/\pi)\log\b$, where the previous assumption $\log a\sim O(\b)$
does not hold anymore. It is then convenient to introduce the new function
$$
         A(x)=a(\l)\;,\quad x=\l-{\g\o\pi}\log{{4\pi\bt}\o\g}     \eqn\resc
$$
Then eqs. \correc\ and \nonlinear\ reduce, in the $\b\to\infty$ limit, to
$$
    L(\b)={2\o{\pi\bt}}\intf dx\,e^{-\pi x/\g}{\rm Im\,}
                                             \log[1+A(x+i0)]  \eqn\cortwo
$$
and$$
        -i\log A(x)= -e^{-\pi x/\g}
        +2\intf dy\,p(x-y)\,{\rm Im\,}\log[1+A(y+i0)]          \eqn\nontwo
$$
Fortunately, it is not necessary to solve this integral equation to calculate
the integral \cortwo. In fact, we can use the following lemma.

Lemma. Assume that $F(x)$ satisfies the nonlinear integral equation
$$
   -i\log F(x)=\varphi(x)+ \int_{x_1}^{x_2}
                  dy\,p(x-y)\,{\rm Im\,}\log[1+F(y+i0)]          \eqn\nlnr
$$
where $\varphi(x)$ is real for real $x$ and $x_1,\,x_2$ are real numbers.
Then the following relation holds
$$\eqalign{
    {\rm Im\,}\int_{x_1}^{x_2}dx\,\varphi^\prime(x)\log[1+F(x+i0)]&=
   \frac12 {\rm
Im\,}\left[\varphi(x_2)\log(1+F_2)-\varphi(x_1)\log(1+F_1)\right]\cr
             &+\frac12 {\rm Re\,}\left[\ell(F_1)-\ell(F_2)\right]\cr} \eqn\lemm
$$
where $F_{1,2}=F(x_{1,2})$ and $\ell$ is a dilogarithm function
$$
      \ell(t)\equiv\int_0^t du\,
            \left[{{\log(1+u)}\o u}-{{\log u}\o{1+u}}\right]    \eqn\dilog
$$
To prove the lemma we consider the relation
$$     \ell(F_2)-\ell(F_1)=
       \int_{x_1}^{x_2}dx\left\{\log[1+F(x)]{d\o{dx}}\log F(x) -
                  \log F(x){d\o{dx}}\log[1+F(x)]\right\}
$$
and then use eq. \nlnr\ and its derivative to substitute $\log F(x)$
and $d\log F(x)/dx$. After taking real part and a little algebra this yields
eq. \lemm\ (related
identities were used in ref. [\aus]).

The integral in eq. \cortwo\ may now be exactly calculated by invoking the
lemma with $F(x)=A(x)$, $\varphi=-\exp(-\pi x/\g)$ and $x_1=-\infty$,
$x_2=+\infty$. We have $A(x_1)=0$, $A(x_2)=1$ and
$$
     2\,{\rm Im}\intf dx\,e^{-\pi x/\g}\log[1+A(x+i0)]=
                               -{\g\o\pi}\ell(1) =-{{\pi\g}\o 6}
$$
Then the free energy for low temperature reads
$$
        f(\b)=E_{XXZ}(\g)-{\g\o{6J\sin\g}}\b^{-2} +o(\b^{-2})   \eqn\lowt
$$
in perfect agreement with refs. [\tak,\devw,\devk]. As we have just shown, both
the
high and the low temperature leading behaviors of the free energy can be
derived without much effort from our non--linear integral equation \nonlinear.
The higher order corrections for high $T$ can be obtained in a very sistematic
way [\noss]. The situation for the $o(\b^{-2})$ terms in the low $T$
expansion \lowt\ is more involved. Qualitatively, however, it is rather easy to
establish, from the asymptotic behaviour of the kernel $p(\l-\mu)$ in
eq. \nonlinear\ and from the $i\pi$ periodicity implied by eq. \discre, that
these higher order terms must be integer powers of $T^{\pi\g}$ and
$T^{\g/(\pi-\g)}$.

Let us now consider the problem of calculating the (properly subtracted) ground
state energy  $E(L)$ of 2D integrable massive models of Quantum Field Theory in
a finite 1--volume $L$. This subject recently received much attention in
connection with Perturbed Conformal Field Theory. In this contest $E(L)$ is
known as ground--state scaling function. Our starting point is the lattice
regularization of such models provided by the light--cone approach to
integrable vertex models [\ddev]. Specializing to the six--vertex model, the
ground state energy, on a ring of length $L$ formed by $N$ sites, takes a form
similar to eq. \ener:
$$
    E_N(L)-E_c(L)= {N\o{\g L}}\,{\rm Im}\intf d\l\,\left[
              {\rm sech\,}\frac\pi\g(\l-\Th)-{\rm sech\,}\frac\pi\g(\l+\Th)
               \right] \log[1+\sub aN(\l+i0)]                     \eqn\enel
$$
where $$
     E_c(L)={{N^2}\o L}\left[-2\pi+\intf d\l\,
           {{\phi(\l+2\Th,\g/2)}\o{\g\cosh\pi\l/\g}}\right]        \eqn\enelc
$$
and $\sub aN(\l+i0)$ obeys an equation like \nonlin\ with the new source term
$$
      \log a_c(\l)=2\arctan{{\sinh\pi\l/\g}\o{\cosh\pi\Th/\g}}  \eqn\sourc
$$
The $L-$dependence in these expressions is hidden in the light--cone parameter
$\Th$, which tends to infinity, in the continuum limit $N\to\infty$, as [\ddev]
$$
                \Th={\g\o\pi}\log{{4N}\o{mL}}                   \eqn\teta
$$
with $m$, the physical mass scale, held fixed. Define now
$$
         \e(\l)=-\lim_{N\to\infty}\log \sub aN(\g\l/\pi)         \eqn\defeps
$$
then from eqs. \nonlin, (42--45) we find
$$
    E(L)\equiv \lim_{N\to\infty}\left[E_N(L)-E_c(L)\right]=
           -m\intf {{d\l}\o \pi}\sinh\l\,{\rm Im\,}\log\left[
            1+e^{-\e(\l+i0)}\right]                              \eqn\gssf
$$
and$$
    \e(\l)=mL\sinh\l+2\intf d\mu\,G(\l-\mu)\,{\rm Im\,}\log\left[
            1+e^{-\e(\mu+i0)}\right]                        \eqn\ntba
$$
where $G(\l)=(\g/\pi)p(\g\l/\pi)$. Thanks to the correspondence between
the light--cone six--vertex model and the sine--Gordon (or Massive Thirring)
model [\ddev], eqs. \gssf\ and \ntba\ give the ground state scaling function
$E(L)$  of the sine--Gordon field theory on a ring of length $L$.(Here
m stands for the fermion mass). This is a
rigorous and simpler alternative to the standard Thermodynamic Bethe Ansatz
[\albz]. Notice that all UV divergent terms are contained in $E_c(L)$
(cft. eq. \enelc). $E_c(L)$ also contains the finite, scaling part
$m^2L{\rm cot^2}\,\pi^2/2\g$ (also found in ref. [\dedev] by
completely different means), which gives the
ground--state energy density in the $L\to\infty$ limit.

Let us now study eqs. \gssf\ and \ntba\ for $mL\gg1$ and $mL\ll1$.
For small values of $mL$ the calculation closely parallels that for low
$T$ in the thermodynamics of the XXZ chain. Shifting $\l\to\l-\log mL$ and
applying the lemma leads to
$$
         E(L)\buildrel{mL\to0}\over=-{\pi\o{6L}}+ O(1)          \eqn\centr
$$
showing the expected value $c=1$ for the UV central charge. The large
$mL$ regime easily follows from eq. \ntba\ by iteration. One finds the behavior
typical of a massive quantum field theory
$$
   E(L)\buildrel{L\to\infty}\over\simeq
                -{{2m}\o\pi}K_1(mL)+ O(e^{-2mL})                \eqn\largel
$$
with $K_1(z)$ the modified Bessel function of order 1.

We would like to remark that, contrary to the traditional thermal Bethe Ansatz
[\tak,\gaud], our approach does not relay on the string hypothesis on the
structure of the solutions of the BA equations characteristic of the
Heisenberg chain. This makes our approach definetly simpler. Most notably,
the whole construction of the thermodynamics no longer depends on whether
$\g/\pi$ is a rational or not, unlike Takahashi approach [\tak]. This
applys equally well to the problem of the ground state scaling function of the
sine--Gordon field theory, since the standard TBA approach requires the
string hypothesis.

Generalizations of the present TBA approach to higher spin chains as
well as to magnetic chains and quantum field theories associated to Lie
algebras other than $A_1$ (nested BA solutions) are relatively
straightforward provided crossing symmetry holds [\noss].

\refout
\bye